# Improved performance in quantum transport calculations: A divide-and-conquer method based on S-matrices

*Mauricio J. Rodríguez, Carlos Ramírez\**

Departamento de Física, Facultad de Ciencias, Universidad Nacional Autónoma de México,

Apartado Postal 70-542, 04510 Ciudad de México, México.

ABSTRACT: We propose a divide-and-conquer algorithm to find recursively the Scattering matrix of general tight-binding structures. The Scattering matrix allows a direct calculation of transport properties in mesoscopic systems by using the Landauer formula. The method is exact, and by analyzing the performance of the algorithm in square, triangular and honeycomb lattices, we show a significant improvement in comparison to other state-of-the-art recursive and non-recursive methods.

**1. Introduction**

Quantum transport is a wide field of study[1–4] for which there exist plenty of recursive methods to study it within the tight-binding approximation.[5–14] These methods are useful to calculate the Green's function, the transfer matrix, or the scattering matrix (S-matrix) of the system, which then can be used to obtain the transmission function $T(E)$ at energy $E$.[5,6] The conductance $G$ of a nanoscale conductor is given in terms of the transmission function, as stated by the celebrated Landauer formula,[5,6]

$$G(E_F) = \frac{e^2}{\pi\hbar} T(E_F) ,\qquad(1)$$

where $E_F$ is the Fermi energy. The transmission function depends strongly on the presence of impurities and defects, the geometry, and the lattice of the systems.

Realistic systems consist of a large number of sites, therefore computational efficiency is an essential part of the recursive methods.[15,16] The transfer matrix can be found by multiplying transfer matrices of slices of the system. This simplicity of operations makes this method preferable, however, it is required that each slice connects to the same number of sites on the left as on the right, which restricts the kind of systems that can be addressed with transfer matrices. On the other hand, Green's function recursive methods are based on the Dyson equation[6,13] that relates the Green's function of a system to one with additional adjacent sites. Each iteration involves the inversion of a $M \times M$ matrix, where $M$ is the number of new sites. In the knitting algoritm[9] sites are added one by one, then this inversion is made on a scalar quantity. The circular slicing algorithm[17] adds groups of sites into rectangular slices which leads to a block tridiagonal Hamiltonian, which is convenient for the recursive method. In addition, an adaptative slicing scheme[15] has been proposed where the first sites to consider are those next to the leads, followed by their first neighbors, then the second neighbors, etc., also leading to a block tridiagonal Hamiltonian. For the case of a square lattice of $L \times L$ sites, the time required to calculate the conductance using these methods scales as $O(L^4)$, being an important improvement in comparison to the direct method, where a $L^2 \times L^2$ matrix is inverted, which scales as $O(L^6)$. However, for tight-binding structures, almost all elements of the Hamiltonian matrix are zero, therefore it is sparse. By using numerical methods suitable for sparse matrices, it has been reported[18] that for $L \sim 10^3$ the computational times scales as $O(L^3)$. But the performance in the memory usage is still better for recursive methods.

In the following, we present a divide-and-conquer algorithm based on the recursive scattering matrix method and we compare its computational efficiency with that of the transfer matrix method for increasingly larger scattering regions

## 2. Recursive Scattering Matrix Method

The recursive S-matrix method (RSMM) allows us to compute the S-matrix of any tight-binding system by using the S-matrices of its subsystems.[19] Building blocks consist of a scattering region connected to auxiliary atomic chains (blue dash-dot-dash lines), as shown in Fig 1. Solutions in the auxiliary chains takes the form of incoming and outgoing waves. The S-matrix relates the amplitude coefficient of such waves, encoding all the effects of the scattering region on its surroundings. The essence of the RSMM is to set the first $N$ incoming (outgoing) amplitude coefficients of a block $A$ equal to the first $N$ outgoing (incoming) amplitude coefficients of a block $B$. In this process, the corresponding auxiliary chains are removed, while the sites connected to these chains in each block overlap, adding their site-energies and hopping integrals. For example, in Fig. 1 the RSMM is used to obtain block $C$, whose S-matrix becomes

$$\mathbf{S}^C = \begin{bmatrix} \mathbf{S}_{22}^A + \mathbf{S}_{21}^A \left(\mathbf{I} - \mathbf{S}_{11}^B \mathbf{S}_{11}^A\right)^{-1} \mathbf{S}_{11}^B \mathbf{S}_{12}^A & \mathbf{S}_{21}^A \left(\mathbf{I} - \mathbf{S}_{11}^B \mathbf{S}_{11}^A\right)^{-1} \mathbf{S}_{12}^B \\ \mathbf{S}_{21}^B \left(\mathbf{I} - \mathbf{S}_{11}^A \mathbf{S}_{11}^B\right)^{-1} \mathbf{S}_{12}^A & \mathbf{S}_{22}^B + \mathbf{S}_{21}^B \left(\mathbf{I} - \mathbf{S}_{11}^A \mathbf{S}_{11}^B\right)^{-1} \mathbf{S}_{11}^A \mathbf{S}_{12}^B \end{bmatrix} \quad (2)$$

being $\mathbf{S}^A_{i,j}$ and $\mathbf{S}^B_{i,j}$ submatrices of the S-matrices of blocks $A$ and $B$, respectively, such that $\mathbf{S}^A_{1,1}$ and $\mathbf{S}^B_{1,1}$ are $N \times N$ matrices. It is worth to mention that the S-matrix of an arbitrary tight-binding structure can be computed starting from the S-matrices of the site and the bond building blocks, illustrated as blocks D and E in Fig. 1, whose S-matrices have a well-known analytic form.[19] The RSMM also allows to model multiterminal systems with general leads.[20] Moreover, an extension of the method gives us the exact order-N Taylor series of the S-matrix and of the transmission function.[21]

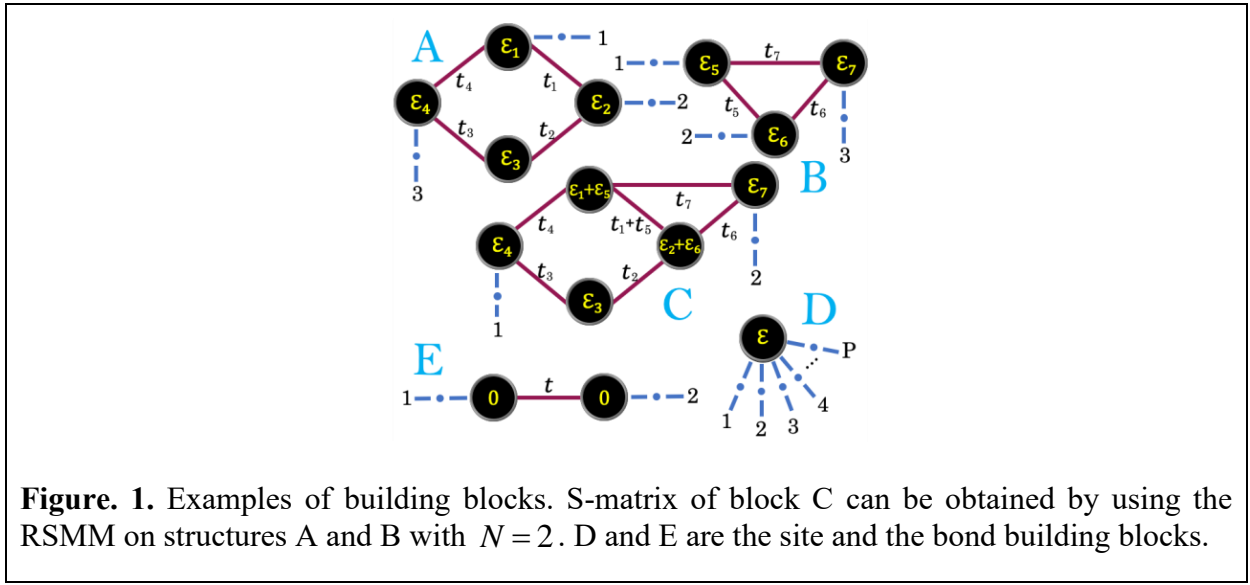

**Figure. 1.** Examples of building blocks. S-matrix of block C can be obtained by using the RSMM on structures A and B with $N=2$. D and E are the site and the bond building blocks.

Following the same ideas than for the knitting algorithm, the RSMM can be run by adding sites one by one. In this scenario, computational time scales as $O(L^4)$ for a square lattice with $L \times L$ sites.[20] However, the RSMM allows other versatile ways to build up the system.

## 3. Divide-and-conquer algorithm

Our method is carried in a divide-and-conquer fashion as illustrated in Fig. 2. It consists in splitting the scattering region into two disconnected subregions, defining a block A. Hopping integrals that connect these subregions form the block B. Observe that by using the RSMM with blocks A and B the original system is recovered. Now, each disconnected region in block A can be calculated independently by splitting again into two disconnected subregions, process that is repeated recursively until one reaches subregions that consist of only one site. Finally, one starts gluing together these regions using the RSMM and following the inverse order that was used to split them.

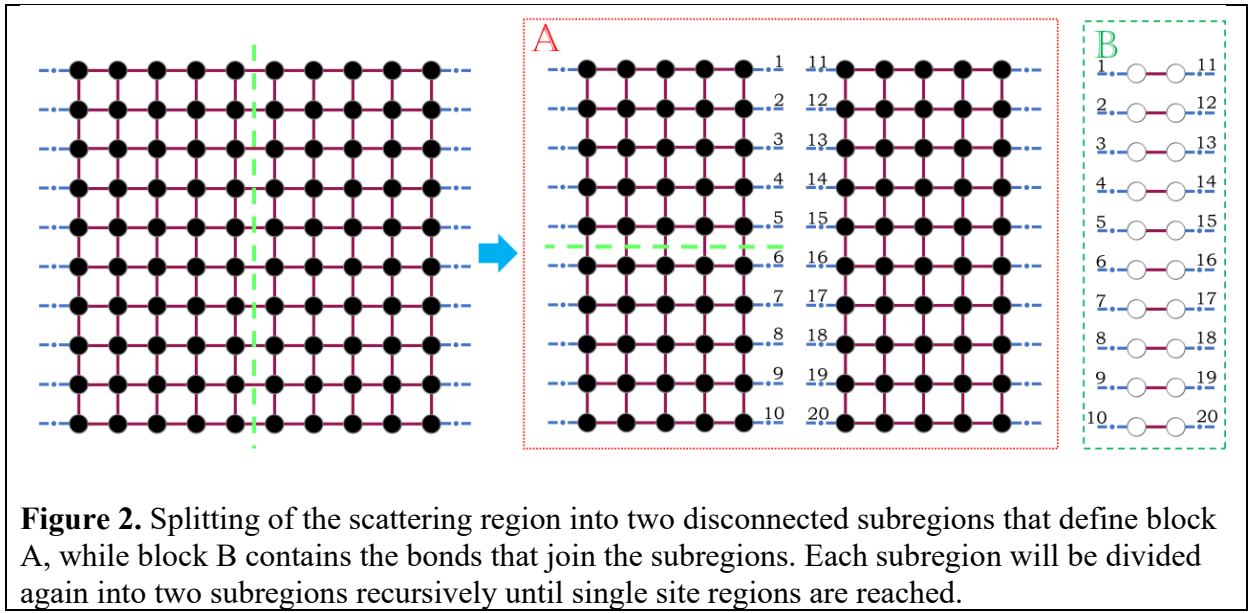

**Figure 2.** Splitting of the scattering region into two disconnected subregions that define block A, while block B contains the bonds that join the subregions. Each subregion will be divided again into two subregions recursively until single site regions are reached.

We make the region's divisions by splitting in half along the horizontal or vertical direction depending on which one is the largest, we do this as a rule of thumb. It is important to clarify that although in Fig. 2 we show a 10×10 square lattice attached to 1D chain leads, this is only an example, and this method is still appliable to an arbitrary shape of the scattering region, lattice, leads, hopping integrals and on-site energies. Furthermore, in case that all on-site energies are $\varepsilon_0$ and all hopping integrals are $t$ then this symmetry allows us to only compute the S-matrix of equivalent regions once, which greatly decreases the computational time needed to calculate the S-matrix of the whole system. However, in this work we study the general case.

## 4. Performance of the divide-and-conquer algorithm

Now, to obtain an analytical estimation of the computational time required by this method, let us consider a square lattice of $L \times L$ sites. To determine the S-matrix of this system, first we have to calculate the S-matrices of four $\frac{L}{2} \times \frac{L}{2}$ subregions, and join them by applying Eq. (2) three times, where multiplications and inversions take a time that scales as $O(L^3)$. Then, the time required to obtain the S-matrix of the $L \times L$ structure, $t(L)$, accomplish the following recursive formula,

$$t(L) = 4t\left(\tfrac{L}{2}\right) + \alpha L^3, \qquad (3)$$

which leads us to,

$$t(L) = 4^N t\left(\tfrac{L}{2^N}\right) + \alpha L^3 \sum_{n=0}^{N-1} \frac{1}{2^n} \approx 4^N t\left(\tfrac{L}{2^N}\right) + 2\alpha L^3. \qquad (4)$$

If $2^N \approx L$ we reach the case were there is only one site per region. Therefore

$$t(L) \approx t_0 L^2 + 2\alpha L^3, \qquad (5)$$

where $t_0$ is the time needed to compute the S-matrix of a site block, while $2\alpha L^3$ is the time consuming during the whole gluing process through the RSMM. By comparing Eqs. (3) and (5), notice that the total time consumed by this gluing process is about twice the time required by the last step, which has no effects on the scaling behavior.

The computational time required to calculate the conductance following the divide-and-conquer algorithm for a square lattice of $L \times L$ sites as a function of $L$ is shown in Fig. 3 with red squares. They fit onto the curve $t(L) = aL^2 + bL^3$ (red line) with $a = 8.509 \times 10^{-5}$ and $b = 3.046 \times 10^{-8}$. Observe that $b \ll a$. To understand the effects of this inequality, let us consider the local scaling behavior by taking $t(L) = cL^d$, where $d(L)$ is calculated by taking the derivative in logscale, obtaining that

$$d = \frac{d \log(t)}{d \log(L)} = 2 + \frac{1}{1 + \frac{a}{bL}}. \qquad (6)$$

Note that for small $L$ we get a local scaling of $L^2$, whereas for very large $L$ it is of $L^3$, and the local scaling makes a soft transition between these two behaviors. For $L \sim 1000$, the local scaling behavior goes as $L^{2.26}$, which represents an improvement in the performance of transport calculations. In terms of the total number of sites, $N = L^2$, this represents an almost $O(N)$ scaling for $N \leq 10^6$ sites, while keeping the calculations exact. On the other hand, for $L \leq 2794$ one gets $d \leq 2.5$, which means that for approximately 8 million total sites our method scales like $L^{2.5}$, staying under the $L^3$ mark of other state-of-the-art methods.

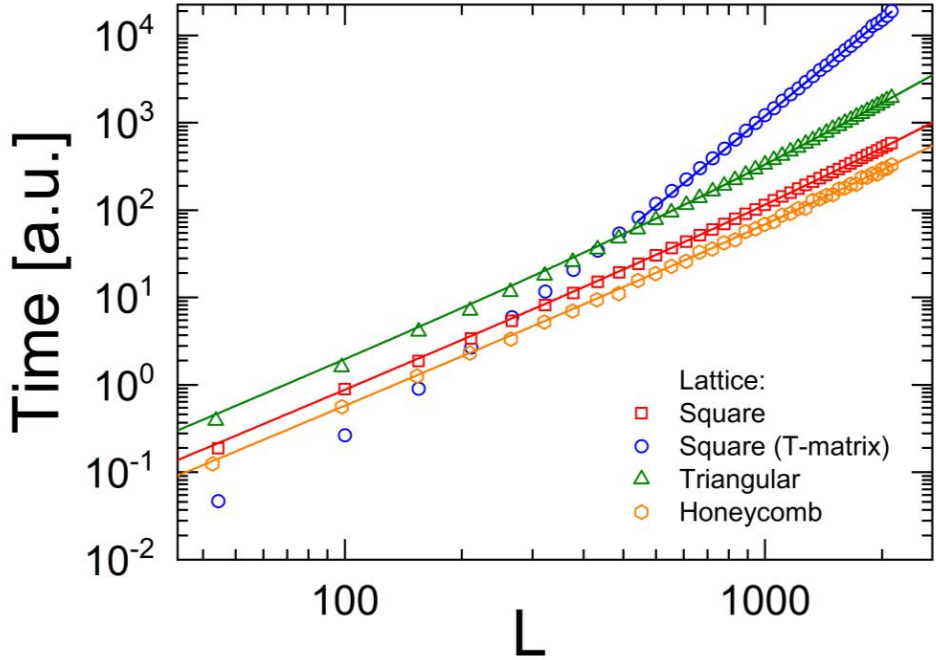

**Figure 3.** Computational time to determine the conductance for a square (red squares), triangular (green triangles) and honeycomb (orange hexagons) lattices using the divide-and-conquer algorithm, and for a square lattice using the transfer matrix method (blue circles). The number of sites in each system is $N = L^2$. Lines correspond to fittings as detailed in the text.

For comparison, Fig. 3 also shows the time to find the conductance by using the transfer matrix method (blue circles). Points for large $L$ are fitted by $\alpha L^{3.971}$ (blue line), with $\alpha = 1.517 \times 10^{-9}$. The very small value of $\alpha$ comes from the simplicity of operations, which makes the transfer matrix method faster than other recursive techniques. However, the divide-and-conquer method has better performance when $L$ is increased. Consequently, we observe that for $L$ greater than $250$, computational time is improved by using the divide-and-conquer algorithm. And even for smaller systems, calculation time between both methods differ by less than one order of magnitude. However, the RSMM is not limited by the restrictions required in the transfer matrix method.

To explore the behavior of the divide-and-conquer algorithm with other lattices, Fig. 3 further shows the computational times for squared areas of triangular (green triangles) and honeycomb lattices (orange hexagons). In these cases, $L$ is related to the number of sites in the area by $L = \sqrt{N}$. Computational times are fitted to $t(L) = aL^2 + bL^3$, where $a = 1.823 \times 10^{-4}$ and $b = 1.513 \times 10^{-7}$ for the triangular lattice (green line), while $a = 5.662 \times 10^{-5}$ and $b = 1.236 \times 10^{-8}$ for the honeycomb lattice (orange line). Notice that coefficients $a$ and $b$ are greater for the triangular lattice and smaller for the honeycomb lattice. This is expected given the nature of the RSMM, because matrices in Eq. (2) grow with the coordination number of the lattice. Beyond this, the scaling behavior is analogous to that of a square lattice, showing improved performance compared to other exact methods for calculating quantum transport properties.

Finally, if we have a square lattice with $L \times W$ sites, where $W \ll L$, the divide-and-conquer algorithm will leads us to calculate $L/W$ S-matrices of blocks of $W \times W$ sites. Then the scaling of the computational time goes as $aLW + bLW^2$. Consequently, for fixed $W$, the divide-and-conquer algorithm is linear in $L$, which is on par with the other recursive methods.

## 5. Conclusions

In summary, we have developed a recursive divide-and-conquer algorithm to compute the S-matrix within the context of the RSMM, that allows us to address multiterminal tight-binding systems with general leads. For a square region with $N = L^2$ sites, the computational time scales as $O(N^{1-1.25})$ for systems with $N \leq 10^6$ sites, an almost linear scaling behavior on $N$. This is an improvement to the $\mathrm{O}(L^4)$ and $\mathrm{O}(L^3)$ marks reported by other recursive and non-recursive state-of-the-art methods. Furthermore, the leading coefficients are small, so that even for small systems, the computational time difference between the transfer matrix method and the divide-and-conquer algorithm is less than an order of magnitude, while for $L > 250$, the latter has the better performance. The divide-and-conquer algorithm can be applied to scattering regions with general geometries and lattices, and its results are exact, since they do not add approximations beyond the tight-binding ones. Moreover, since each subregion is independent, their S-matrices can be calculated in parallel, which can make the algorithm even faster. We expect that this paper inspires a new generation of high-performance algorithms to study quantum transport properties in real-size systems.

## AUTHOR INFORMATION

**Corresponding Author**

*E-mail address: carlos@ciencias.unam.mx

**Author Contributions**

The manuscript was written through contributions of all authors. All authors have given approval to the final version of the manuscript. Authors contributed equally.

## ACKNOWLEDGMENT

This work has been supported by UNAM-DGAPA-PAPIIT IN109022. Computations were performed at Miztli under project LANCAD-UNAM-DGTIC-329.